\documentclass[aps,twocolumn,showpacs,preprintnumbers,floats]{revtex4}\def\@cite#1#2{\textsuperscript{[{#1\if@tempswa , #2\fi}]}}

\usepackage{graphicx}
\usepackage{amsmath}
\usepackage{amsfonts}
\usepackage{amssymb}
\usepackage{color}
\usepackage{subfigure}
\usepackage{epsfig}
\usepackage{morefloats}
\usepackage{multirow}
\usepackage{graphicx,booktabs}
\usepackage{mathrsfs}
\usepackage{txfonts}
\usepackage{indentfirst}
\usepackage{graphicx,booktabs}

\usepackage{longtable,lscape}

\newcommand{\vsig}{\mbox{\boldmath$\sigma$\unboldmath}}

\usepackage{comment}

\begin{document}

\title{Interpretation of the newly observed $\Lambda_b(6146)$ and $\Lambda_b(6152)$ states in a chiral quark model}
\author{
Kai-Lei Wang$^{1}$~\footnote {E-mail: wangkaileicz@foxmail.com}, Qi-Fang L\"{u}$^{2,3}$~\footnote {E-mail: lvqifang@hunnu.edu.cn}, and Xian-Hui Zhong$^{2,3}$~\footnote {E-mail: zhongxh@hunnu.edu.cn}}

\affiliation{ 1) Department
of Electronic Information and Physics, Changzhi University, Changzhi, Shanxi,046011,China}

\affiliation{ 2) Department
of Physics, Hunan Normal University, and Key Laboratory of
Low-Dimensional Quantum Structures and Quantum Control of Ministry
of Education, Changsha 410081, China }

\affiliation{ 3) Synergetic Innovation Center for Quantum Effects and Applications (SICQEA),
Hunan Normal University, Changsha 410081, China}

\begin{abstract}
The strong decays of the low-lying $\lambda$-mode $\Lambda_b(1D,2S)$ and $\Sigma_b(2S)$ states are studied in a chiral quark model. We find that: (i) the newly observed $\Lambda_b(6146/6152)$ resonances in the $\Lambda_b\pi^+\pi^-$ spectrum  by the LHCb Collaboration
might be explained with the $\lambda$-mode $\Lambda_b(1D)$ states in the quark model. It should be emphasized that whether the structure in the $\Lambda_b\pi^+\pi^-$ spectrum correspond to two states or one state should be further clarified with more observations in future experiments. (ii) The $\Lambda_b(2S)$ $|J^P = \frac{1}{2}^+,0\rangle$ state mainly decays into $\Sigma_b^*\pi$ channel, which may be an ideal channel for searching for this $\Lambda_b(2S)$ state in future experiments. (iii) The $\Sigma_b(2S)$ $|J^P = \frac{1}{2}^+,1 \rangle$ and $|J^P = \frac{3}{2}^+,1 \rangle$ dominantly decay into $\Lambda_b\pi$ with $\Gamma \simeq 3.82$ MeV and $\Gamma \simeq 4.72$ MeV, respectively.
\end{abstract}

\maketitle

\section{Introduction}{\label{introduction}}

Understanding the heavy baryon spectrum and searching for the missing heavy baryons and new exotic states are interesting topics in hadronic physics~\cite{Tanabashi:2018oca,Chen:2016spr,Cheng:2015iom,Crede:2013sze,Klempt:2009pi}. The LHC facility provides good opportunities for us to establish the heavy baryon spectra. In 2017, five extremely narrow $\Omega_c$ states, $\Omega_c(3000)$, $\Omega_c(3050)$, $\Omega_c(3066)$, $\Omega_c(3090)$, and $\Omega_c(3119)$,
were observed  in the $\Xi_c^{+}K^-$ channel by the LHCb Collaboration~\cite{Aaij:2017nav}.
Most of them may be interpreted as the $P$-wave excited states of $\Omega_c$~\cite{Wang:2017hej,Cheng:2017ove,
Chen:2017gnu,Padmanath:2017lng,Wang:2017vnc, Karliner:2017kfm,Chen:2017sci,Agaev:2017lip,Wang:2017zjw}.
In 2018, the LHCb collaboration announced the observation of a
new bottom baryon $\Xi_b(6227)^-$ in both
$\Lambda_b K^-$ and $\Xi_b^0\pi^-$ decay modes~\cite{Aaij:2018yqz} and found two new resonances $\Sigma_b(6097)^{\pm}$
in the $\Lambda_b \pi^{\pm}$ channels~\cite{Aaij:2018tnn}. Both the $\Sigma_b(6097)^{\pm}$ and $\Xi_b(6227)^-$ states favor $P$ wave states with spin-parity $J^P = 3/2^-$ or $J^P = 5/2^-$\cite{Yang:2018lzg,Aliev:2018vye,Cui:2019dzj,Wang:2018fjm,Chen:2018orb,Chen:2018vuc}.
Recently, the LHCb collaboration announced the observation of two
new bottom baryons $\Lambda_b(6146)$ and $\Lambda_b(6152)$ in the
$\Lambda_b\pi^+\pi^-$ decay mode~\cite{Aaij:2019amv}.  The measured masses and widths of the $\Lambda_b(6146)$ and $\Lambda_b(6152)$ are presented as follows,
\begin{eqnarray}
m[\Lambda_b(6146)] = 6146.17\pm0.33\pm0.22\pm0.16~\rm{MeV},
\end{eqnarray}
\begin{eqnarray}
\Gamma[\Lambda_b(6146)] = 2.9\pm1.3\pm0.3~\rm{MeV},
\end{eqnarray}
\begin{eqnarray}
m[\Lambda_b(6152)] = 6152.51\pm0.26\pm0.22\pm0.16~\rm{MeV},
\end{eqnarray}
\begin{eqnarray}
\Gamma[\Lambda_b(6152)] = 2.1\pm0.8\pm0.3~\rm{MeV}.
\end{eqnarray}
Considering their masses and small splitting, the LHCb Collaboration suggested that both of them may be clarified into the doublet of $\Lambda_b(1D)$ states in the quark model. It should be noticed that current experimental information does not exclude the two new resonances as the $I=0$ $\Sigma_b$ states, which has to be considered in the following discussions.

\begin{table*}[htp]
\begin{center}
\caption{\label{sp1}  Mass spectra of the $\Lambda_b$ up to $D$ wave and the $\Sigma_b$ up to $2S$ wave from
various quark models~\cite{Ebert:2011kk,Yoshida:2015tia,Roberts:2007ni,Capstick:1986bm,Chen:2014nyo} compared
with the data from the Particle Data Group~\cite{Tanabashi:2018oca}. The units are in MeV.}
\begin{tabular}{c|ccccccccccccccccccccccccccccccccccccccccccccc}\hline\hline
&\multicolumn{6}{c}{$\underline{~~~~~~~~~~~~~~~~~~~~~~~~~~~~~~~~~~~~~~~~~~~~~~~~~~~~~~~~~~~~~~~~~~~~~~~~~~~~~~~~\Lambda_b~~~~~~~~~~~~~~~~~~~~~~~~~~~~~~~~~~~~~~~~~~~~~~~~~~~~~~~~~~~~~~~~~~~~~~~}$}\\
State            &~~~~\cite{Ebert:2011kk} &~~~~~~~~~~~~\cite{Yoshida:2015tia} &~~~~~~~~~~~~\cite{Roberts:2007ni} &~~~~~~~~~~~~\cite{Capstick:1986bm}   &~~~~~~~~~~~~\cite{Chen:2014nyo} &~~~~~~~~~~~~PDG\cite{Tanabashi:2018oca}\\ \hline
$|J^P=\frac{1}{2}^+,0 \rangle$        &~~~~5620      &~~~~~~~~~~~~5618   &~~~~~~~~~~~~5612  &~~~~~~~~~~~~5585    &~~~~~~~~~~~~5619 &~~~~~~~~~~~~5620 \\
$|J^P=\frac{1}{2}^-,1\rangle$         &~~~~5930      &~~~~~~~~~~~~5938   &~~~~~~~~~~~~5939  &~~~~~~~~~~~~5912    &~~~~~~~~~~~~5911 &~~~~~~~~~~~~5912\\
$|J^P=\frac{3}{2}^-,1\rangle$         &~~~~5942      &~~~~~~~~~~~~5939   &~~~~~~~~~~~~5941  &~~~~~~~~~~~~5920    &~~~~~~~~~~~~5920 &~~~~~~~~~~~~5920\\
$|J^P=\frac{1}{2}^+,0\rangle$         &~~~~6089      &~~~~~~~~~~~~6153   &~~~~~~~~~~~~6107  &~~~~~~~~~~~~6045       &~~~~~~~~~~~~$\cdot\cdot\cdot$     &~~~~~~~~~~~~? \\
$|J^P=\frac{3}{2}^+,2\rangle$         &~~~~6190      &~~~~~~~~~~~~6211   &~~~~~~~~~~~~6181  &~~~~~~~~~~~~6145    &~~~~~~~~~~~~6147 &~~~~~~~~~~~~ ?\\
$|J^P=\frac{5}{2}^+,2\rangle$         &~~~~6196      &~~~~~~~~~~~~6212   &~~~~~~~~~~~~6183  &~~~~~~~~~~~~6165    &~~~~~~~~~~~~6153 &~~~~~~~~~~~~?\\

\hline\hline
&\multicolumn{6}{c}{$\underline{~~~~~~~~~~~~~~~~~~~~~~~~~~~~~~~~~~~~~~~~~~~~~~~~~~~~~~~~~~~~~~~~~~~~~~~~~~~~~~~~\Sigma_b~~~~~~~~~~~~~~~~~~~~~~~~~~~~~~~~~~~~~~~~~~~~~~~~~~~~~~~~~~~~~~~~~~~~~~~}$}\\
State        &~~~~\cite{Ebert:2011kk}       &~~~~~~~~~~~~ \cite{Yoshida:2015tia}  &~~~~~~~~~~~~\cite{Roberts:2007ni} &~~~~~~~~~~~~\cite{Capstick:1986bm}  &~~~~~~~~~~~~ &~~~~~~~~~~~~PDG~\cite{Tanabashi:2018oca}\\ \hline
$|J^P=\frac{1}{2}^+,1 \rangle$          &~~~~5808       &~~~~~~~~~~~~5823     &~~~~~~~~~~~~5833   &~~~~~~~~~~~~5795  &~~~~~~~~~~~~ &~~~~~~~~~~~~5811 \\
$|J^P=\frac{3}{2}^+, 1\rangle$          &~~~~5834       &~~~~~~~~~~~~5845     &~~~~~~~~~~~~5858   &~~~~~~~~~~~~5805  &~~~~~~~~~~~~      &~~~~~~~~~~~~5832 \\
$|J^P=\frac{1}{2}^-,0\rangle$           &~~~~6101       &~~~~~~~~~~~~6127     &~~~~~~~~~~~~6099   &~~~~~~~~~~~~6070  &~~~~~~~~~~~~      &~~~~~~~~~~~~?\\
$|J^P=\frac{1}{2}^-,1\rangle$           &~~~~6095       &~~~~~~~~~~~~6135     &~~~~~~~~~~~~6106   &~~~~~~~~~~~~6070  &~~~~~~~~~~~~        &~~~~~~~~~~~~ ?\\
$|J^P=\frac{3}{2}^-,1\rangle$           &~~~~6096       &~~~~~~~~~~~~6132     &~~~~~~~~~~~~6101   &~~~~~~~~~~~~6070  &~~~~~~~~~~~~            &~~~~~~~~~~~~?\\
$|J^P=\frac{3}{2}^-,2\rangle$           &~~~~6087       &~~~~~~~~~~~~6141     &~~~~~~~~~~~~6105   &~~~~~~~~~~~~6085  &~~~~~~~~~~~~           &~~~~~~~~~~~~ ?\\
$|J^P=\frac{5}{2}^-,2\rangle$           &~~~~6084       &~~~~~~~~~~~~6144     &~~~~~~~~~~~~6172   &~~~~~~~~~~~~6090  &~~~~~~~~~~~~            &~~~~~~~~~~~~?\\
$|J^P=\frac{1}{2}^+,1\rangle$           &~~~~6213       &~~~~~~~~~~~~6343     &~~~~~~~~~~~~6294   &~~~~~~~~~~~~6200  &~~~~~~~~~~~~          &~~~~~~~~~~~~?\\
$|J^P=\frac{3}{2}^+,1\rangle$           &~~~~6226       &~~~~~~~~~~~~         &~~~~~~~~~~~~6308   &~~~~~~~~~~~~6250  &~~~~~~~~~~~~           &~~~~~~~~~~~~ ?\\

\hline\hline
\end{tabular}
\end{center}
\end{table*}

Since the $\Lambda_b$ and $\Sigma_b$ baryons contains a heavy bottom quark and two light $u/d$ quarks, the low-lying internal excitations favor  excitations of the so-called $\lambda$-mode where the orbital excitation lies between the light
quarks and the heavy quark in a Jacobi coordinate. In the $\Lambda_b$ family,  the ground state $\Lambda_b$, and two low-lying orbitally excited states $\Lambda_b(5912)~1/2^-$ and $\Lambda_b(5920)~3/2^-$,  have been well established \cite{Capstick:1986bm,Ebert:2007nw,Roberts:2007ni,Ebert:2011kk,Chen:2014nyo,Garcilazo:2007eh,Karliner:2015ema,
Mao:2015gya,Thakkar:2016dna,Wang:2017kfr}.  According to the mass spectrum predicted in various models~\cite{Ebert:2007nw,Roberts:2007ni,Ebert:2011kk,Chen:2014nyo,Mao:2015gya,Capstick:2000qj,
Ebert:2005xj,Yoshida:2015tia,Valcarce:2008dr,Karliner:2008sv,Chen:2016phw}, the two newly observed resonances $\Lambda_b(6146)$ and $\Lambda_b(6152)$, lie in the mass region of the $\lambda$-mode $\Lambda_b(1D,2S)$ states and the neutral
excited $\Sigma_b(2S)$ states.

To identify the $\Lambda_b(6146)$ and $\Lambda_b(6152)$ resonances in the quark model, their strong decay behaviors were studied
with the $^3P_0$ model in Ref.~\cite{Liang:2019aag}.
The results indicated that the $\Lambda_b(6146)$ and $\Lambda_b(6152)$ can be assigned as the $\Lambda_b(1D)$ doublet with $J^P=5/2^+$ and $J^P=3/2^+$, respectively. In our previous works~\cite{Wang:2017kfr,Wang:2018fjm},  we have studied the strong decays of the singly bottom baryons $\Sigma_b(1P)$ within the chiral quark model.  We find that the $\Sigma_b(6097)$ favors the light
spin $j = 2$ states with spin-parity numbers $J^P=3/2^-$ or $J^P=5/2^-$. If $\Sigma_b(6097)$ corresponds to the $|J^P=\frac{3}{2}^-,2\rangle$ or  $|J^P=\frac{5}{2}^-,2\rangle$ state, the typical mass for the  $\Sigma_b(1P)$ states should be around 6090 MeV since the mass splitting between them is within 10 MeV according to quark model predictions~\cite{Ebert:2011kk}. Thus, one should exclude the $\Sigma_b(6146)$ and $\Sigma_b(6152)$ resonances as the $\Sigma_b(1P)$ states for the masses of $\Sigma_b(6146)$ and $\Sigma_b(6152)$ are obviously larger than the quark model predictions.

In previous works, the decays of the $P$- and $D$-wave singly heavy baryons have been studied within the chiral quark model~\cite{Yao:2018jmc,Wang:2017kfr}. Since the newly observed $\Lambda_b(6146)$ and $\Lambda_b(6152)$ resonances may favor the $D$-wave $\Lambda_b$ states, to confirm this assignment, in present work we revisit the strong decays of the $D$-wave $\Lambda_b$ states by adopting the measured masses. Moreover, the $2S$-wave $\Lambda_b$ and $\Sigma_b$ states were not investigated in the frame work of chiral quark model. Hence, in this work, as a supplement of Refs.~\cite{Yao:2018jmc,Wang:2017kfr}, we employ the chiral quark model to study the strong decays with emission of one light
pseudoscalar meson for the low-lying $\Lambda_b(1D,2S)$ and $\Sigma_b(2S)$ states.
Combining the masses, total widths and decay modes, our results suggest that the $\Lambda_b(6146)$ and $\Lambda_b(6152)$ states may favor the $\Lambda_b(1D)$ $|J^P = \frac{5}{2}^+,2 \rangle$ and $|J^P = \frac{3}{2}^+,2 \rangle$ states, respectively.

This paper is organized as follows. The spectrum and notations are presented in Sec.~\ref{spectrum}. The chiral quark model are briefly introduced in Sec.~\ref{model}. The  strong decays of the low-lying $\Lambda_b(1D,2S)$ and $\Sigma_b(2S)$ states are estimated in Sec.~\ref{results}. A short summary is presented in the last section.

\section{SPECTROSCOPY }\label{spectrum}

\begin{table}[htp]
\begin{center}
\caption{\label{JJ1}  The classifications of the low-lying $1P$ and $2S$-wave states belonging to $\mathbf{6}_F$ in the $j-j$ coupling scheme.}
\begin{tabular}{p{2.0cm}p{0.5cm}p{0.5cm}p{0.5cm}p{0.5cm}p{0.5cm}p{0.5cm}p{0.5cm}p{0.5cm}p{0.5cm}p{0.5cm}p{0.5cm} |}\hline\hline
$|J^P,j \rangle$      &$J^P$            &$j$  &$n_{\rho}$ &$\ell_{\rho}$  &$n_{\lambda}$  &$\ell_{\lambda}$    &$L$   &$s_{\rho}$    &$s_{Q}$                     \\ \hline
$|J^P=\frac{1}{2}^-,0\rangle$   &$\frac{1}{2}^-$  & 0    & 0    & 0      & 0          &1          &1       &1         &$\frac{1}{2}$         \\
$|J^P=\frac{1}{2}^-,1\rangle$   &$\frac{3}{2}^-$  & 1    & 0    & 0      & 0          &1          &1       &1         &$\frac{1}{2}$              \\
$|J^P=\frac{3}{2}^-,1\rangle$   &$\frac{1}{2}^-$  & 1    & 0    & 0      & 0          &1          &1       &1         &$\frac{1}{2}$           \\
$|J^P=\frac{3}{2}^-,2\rangle$   &$\frac{3}{2}^-$  & 2    & 0    & 0      & 0          &1          &1       &1         &$\frac{1}{2}$            \\
$|J^P=\frac{5}{2}^-,2\rangle$   &$\frac{5}{2}^-$  & 2    & 0    & 0      & 0          &1          &1       &1         &$\frac{1}{2}$            \\
$|J^P=\frac{1}{2}^+,1\rangle$   &$\frac{1}{2}^+$  & 1    & 0    & 0      & 1          &0          &0       &1         &$\frac{1}{2}$            \\
$|J^P=\frac{3}{2}^+,1\rangle$   &$\frac{3}{2}^+$  & 1    & 0    & 0      & 1          &0          &0       &1         &$\frac{1}{2}$            \\

\hline\hline
\end{tabular}
\end{center}
\end{table}

\begin{table}[htp]
\begin{center}
\caption{\label{JJ2}  The classifications of the low-lying $1P$, $1D$ and $2S$-wave states belonging to $\mathbf{\bar{3}}_F$ in the $j-j$ coupling scheme.}
\begin{tabular}{p{2.0cm}p{0.5cm}p{0.5cm}p{0.5cm}p{0.5cm}p{0.5cm}p{0.5cm}p{0.5cm}p{0.5cm}p{0.5cm} }\hline\hline
$|J^P,j \rangle$      &$J^P$            &$j$  &$n_{\rho}$ &$\ell_{\rho}$  &$n_{\lambda}$  &$\ell_{\lambda}$    &$L$   &$s_{\rho}$    &$s_{Q}$                     \\ \hline
$|J^P=\frac{1}{2}^-,1\rangle$   &$\frac{1}{2}^-$  & 0    &0    & 0    &0            &1          &1       &1         &$\frac{1}{2}$         \\
$|J^P=\frac{3}{2}^-,1\rangle$   &$\frac{3}{2}^-$  & 1    &0    & 0    &0            &1          &1       &0         &$\frac{1}{2}$ \\
$|J^P=\frac{1}{2}^+,0\rangle$   &$\frac{1}{2}^+$  & 0    &0    & 0    &1            &0          &0       &0         &$\frac{1}{2}$\\
$|J^P=\frac{3}{2}^+,2\rangle$   &$\frac{3}{2}^+$  & 2    &0    & 0    &0            &2          &2       &0         &$\frac{1}{2}$\\
$|J^P=\frac{5}{2}^+,2\rangle$   &$\frac{3}{2}^+$  & 2    &0    & 0    &0            &2          &2       &0         &$\frac{1}{2}$\\
\hline\hline
\end{tabular}
\end{center}
\end{table}

The heavy baryon containing a heavy quark violates
the SU(4) symmetry. However, the SU(3) symmetry between the other two
light quarks ($u$, $d$, or $s$) is approximately kept. According to
the symmetry, the heavy baryons containing a single heavy quark
belong to two different SU(3) flavor representations: the symmetric sextet $\mathbf{6}_F$ and
antisymmetric antitriplet $\bar{\mathbf{3}}_F$.
In the singly bottom baryons, $\Lambda_b$ and $\Xi_b^{0,-}$ belonging to $\bar{\mathbf{3}}_F$ representation and the $\Sigma_b^{-,0,+}$, $\Xi_b'^{0,-}$ and $\Omega_b$ form a $\mathbf{6}_F$ representation.

It should be pointed out that  the quark model states that we obtain for the
baryons containing a single heavy quark respect the dictates of the heavy quark effective theory (HQET).
In the heavy quark effective
theory description,
the total angular momentum of the two light quarks $\mathbf{j}=\mathbf{L}+\mathbf{s}_{\rho}$ is conserved and coupled to the spin of a heavy quark with spin $\mathbf{s}_Q=1/2$~\cite{Cheng:2015iom,Roberts:2007ni}. The total angular
momentum can take the values $\mathbf{J}=\mathbf{j}+\mathbf{s}_Q$.
In the heavy-quark
symmetry limit, the quark model states may more favor the
$j-j$ coupling scheme
\begin{eqnarray}
|J^P,j\rangle = |\{[(\ell_\rho \ell_\lambda)_L s_{\rho}]_js_Q\}_{J^P}\rangle.
\end{eqnarray}
The $\Sigma_b$ $1P$- and $2S$-wave states belonging to $\mathbf{6}_F$ and the $\Lambda_b$ $1P$-, $2S$- and $1D$-wave states belonging to $\bar{\mathbf{3}}_F$ in the $j-j$ coupling scheme and
their corresponding quantum numbers have been collected in Table~\ref{JJ1} and \ref{JJ2}.

\section{ Strong decay  }\label{model}

In this work,  strong decays of the  singly heavy baryons with emission of one light pseudoscalar meson, i.e. $\pi$, $K$ and $\eta$, are studied within chiral quark model~\cite{Manohar:1983md}.
In this model, the light pseudoscalar mesons
are treated as point-like Goldstone boson.
This model has been successfully applied to study the strong decays of heavy-light mesons
and strange and singly heavy baryons~\cite{Zhong:2008kd,Zhong:2010vq,Zhong:2009sk,
Zhong:2007gp,Liu:2012sj,Xiao:2013xi,Nagahiro:2016nsx,Wang:2017hej,Xiao:2014ura,Xiao:2017udy,Yao:2018jmc}.
The effective quark-pseudoscalar-meson interactions  at low energies
can be described by the simple chiral Lagrangian~\cite{Zhong:2008kd,Zhong:2010vq,Zhong:2009sk,
Zhong:2007gp,Liu:2012sj,Xiao:2013xi,Wang:2017hej,Xiao:2014ura}:
\begin{equation}\label{coup}
H_{m}=\sum_j
\frac{1}{f_m}\bar{\psi}_j\gamma^{j}_{\mu}\gamma^{j}_{5}\psi_j\partial^{\mu}\phi_m,
\end{equation}
where $\psi_j$ represents the $j$th quark field in the hadron;
$\phi_m$ is the pseudoscalar meson field, $f_m$ is the
pseudoscalar meson decay constant.

To match the nonrelativistic harmonic oscillator wave functions
in this work, one should adopt the quark-pseudoscalar-meson interactions in
the nonrelativistic form~\cite{Zhong:2008kd,Zhong:2010vq,Zhong:2009sk,
Zhong:2007gp,Liu:2012sj,Xiao:2013xi,Wang:2017hej,Xiao:2014ura}:
\begin{eqnarray}\label{ccpk}
H_{m}^{nr}=\sum_j\left[ \mathcal{G} \vsig_j \cdot \textbf{q}
+h \vsig_j\cdot \textbf{p}_j\right]I_j
e^{-i\mathbf{q}\cdot \mathbf{r}_j},
\end{eqnarray}
with $\mathcal{G}\equiv -(1+\frac{\omega_m}{E_f+M_f})$ and $h\equiv \frac{\omega_m}{2\mu_q}$.
In the above equation, $\omega_m$ and $\mathbf{q}$ are the energy and three momenta of the emitted light meson,
respectively;  $\mu_q$ stand for a reduced mass given by $1/\mu_q=1/m_j+1/m'_j$ with
$m_j$ and $m'_j$ for the masses of the $j$th quark in the initial and
final hadrons, respectively; $\vsig_j$ and $\textbf{p}_j$ are the
Pauli spin vector and internal momentum operator for the $j$th quark of the initial hadron;
and $I_j$ is the isospin operator associated with the pseudoscalar meson.

For a light pseudoscalar meson emission in a strong decay process,
the partial decay width can be calculated with~\cite{Zhong:2008kd, Zhong:2007gp}
\begin{equation}\label{dww}
\Gamma_m=\left(\frac{\delta}{f_m}\right)^2\frac{(E_f+M_f)|\mathbf{q}|}{4\pi
M_i(2J_i+1)} \sum_{J_{fz},J_{iz}}|\mathcal{M}_{J_{fz},J_{iz}}|^2 ,
\end{equation}
where $\mathcal{M}_{J_{fz},J_{iz}}$  corresponds to
the strong  amplitudes.
The quantum numbers $J_{iz}$ and $J_{fz}$ stand for the third components of the total
angular momenta of the initial and final heavy baryons,
respectively. $M_i$ is the mass of the initial heavy baryon. $E_f$ and $M_f$ are the energy and mass of the final heavy baryon.
$\delta$ as a global parameter accounts for the
strength of the quark-meson couplings.  It has been determined in our previous work of the strong
decays of the charmed baryons and heavy-light mesons
\cite{Zhong:2007gp,Zhong:2008kd}. Here, we fix its value the same as
that in Refs.~\cite{Zhong:2008kd,Zhong:2007gp}, i.e. $\delta=0.557$.

In the calculation, the standard quark model parameters are
adopted. Namely, we set $m_u=m_d=330$ MeV and $m_b=5000$ MeV for the constituent quark masses.
The harmonic oscillator parameter $\alpha_{\rho}$ in the wave function
$\psi^n_{lm}=R_{nl}Y_{lm}$ for $uu/ud/dd$ diquark systems is taken as $\alpha_{\rho}=400$ MeV. Another harmonic oscillator parameter $\alpha_{\lambda}$ can
be related to $\alpha_{\rho}$ with the relation $\alpha^2_\lambda=\sqrt{3m_b/(2m+m_b)}\alpha^2_\rho$. The decay
constant for $\pi$  meson is taken as $f_\pi = 132$ MeV. The masses of the well-established hadrons used in the calculations are taken from the Particle Data Group (PDG)~\cite{Tanabashi:2018oca}, and the masses of the undiscovered initial states adopt  from the predictions in Refs.~\cite{Ebert:2011kk,Capstick:1986bm}.

\section{Results and discussions}\label{results}

\begin{table}[htb]
\begin{center}
\caption{ \label{Lamb1D} Partial widths (MeV) of strong decays for the $\lambda$-mode $1D$-wave   $\Lambda_b$ baryons.}
%\footnotesize
\begin{tabular}{c|ccccc}
\hline\hline
 \multirow{2}{*}{Decay mode}        &\multicolumn{2}{c}{$\underline{~~~~~~|J^P = \frac{3}{2}^+,2 \rangle~~~~~~}$} &\multicolumn{2}{c}{$\underline{~~~~~~|J^P = \frac{5}{2}^+,2 \rangle~~~~~~}$} \\
             &~~ $\Lambda_b(6146)$         &$\Lambda_b(6152)$    &~~ $\Lambda_b(6146)$         &$\Lambda_b(6152)$ \\ \hline
  $\Sigma_b\pi$                                &4.41    &4.67     &0.64   &0.73  \\
 $\Sigma_b^*\pi$                               &1.26    &1.41     &4.26   &4.60  \\
Sum                                            &5.67    &6.08     &4.90   &5.33 \\\hline
\hline
\end{tabular}
\end{center}
\end{table}

%Our results are listed in Table.\ref{Lamb2S}

\subsection{$\Lambda_b(1D)$}

In the $\Lambda_b$ family,  there are two $\lambda$-mode $1D$-wave
excitations $|J^P = \frac{3}{2}^+,2 \rangle$
and $|J^P = \frac{5}{2}^+,2 \rangle$ according to the classification of quark models.
The masses for the $\lambda$-mode $1D$-wave $\Lambda_b$
excitations are predicted to be $\sim 6.1$-$6.2$ GeV (see Table~\ref{sp1}).
The measured masses of the $\Lambda_b(6146)$ and $\Lambda_b(6152)$ indicate that they may be  good candidates of the $\lambda$-mode $1D$-wave excitations.

Our calculations are presented in Table~\ref{Lamb1D}. The $\Lambda_b(6146)$ resonance is most likely to be the $J^P = 5/2^+$ $|J^P = \frac{5}{2}^+,2 \rangle$ state. Assigning the $\Lambda_b(6146)$ as $|J^P = \frac{5}{2}^+,2 \rangle$,  one can find that it has a narrow width of $\sim$ 5 MeV, and dominantly decays into $\Sigma_b^*\pi$. The partial widths into $\Sigma_b\pi$ and $\Sigma_b^*\pi$ channels are predicted to be
\begin{eqnarray}\label{1DL}
\Gamma[\Lambda_b(6146) \rightarrow \Sigma_b\pi]\simeq 0.64 \ \mathrm{MeV}, \\
\Gamma[\Lambda_b(6146) \rightarrow \Sigma_b^*\pi]\simeq 4.26 \ \mathrm{MeV}.
\end{eqnarray}
Both the decay width and decay mode are consistent with the observations of $\Lambda_b(6146)$  considering the model uncertainties. This conclusion is consistent with that of the $^3P_0$ model~\cite{Liang:2019aag}.
To further confirm the nature of the $\Lambda_b(6146)$ resonance,  the partial width ratio between $\Sigma_b\pi$ and $\Sigma_b^*\pi$ channels,
\begin{eqnarray}\label{aa}
\frac{\Gamma(\Sigma_b\pi)}{\Gamma(\Sigma_b^*\pi)}\simeq 0.15,
\end{eqnarray}
is worth to observing in future experiments.

On the other hand, the  $\Lambda_b(6152)$ resonance most likely corresponds to the $\lambda$-mode $1D$-wave $\Lambda_b$ excitation $|J^P = \frac{3}{2}^+,2 \rangle$. When we assign the $\Lambda_b(6152)$ as $|J^P = \frac{3}{2}^+,2 \rangle$, it has a narrow width of $\sim$ 6 MeV, and dominantly decays into $\Sigma_b\pi$ and $\Sigma_b^*\pi$. We predicted partial widths
\begin{eqnarray}\label{1DL}
\Gamma[\Lambda_b(6152) \rightarrow \Sigma_b\pi]\simeq 4.67 \ \mathrm{MeV}, \\
\Gamma[\Lambda_b(6152) \rightarrow \Sigma_b^*\pi]\simeq 1.41 \ \mathrm{MeV},
\end{eqnarray}
which are in agreement with the the $^3P_0$ model predictions $\Gamma[\Lambda_b(6152) \rightarrow \Sigma_b\pi]\simeq 5.31 $ MeV and $\Gamma[\Lambda_b(6152) \rightarrow \Sigma_b\pi]\simeq 0.87 $ MeV~\cite{Liang:2019aag}. The partial width ratio between $\Sigma_b\pi$ and $\Sigma_b^*\pi$ channels is predicted to be
\begin{eqnarray}\label{aa}
\frac{\Gamma(\Sigma_b\pi)}{\Gamma(\Sigma_b^*\pi)}\simeq 3.3.
\end{eqnarray}
This ratio might be crucial to test the nature of $\Lambda_b(6152)$, which is suggested to be measured in future experiments.

Finally, it should be mentioned that the $\Lambda_c(2860)3/2^+$ and $\Lambda_c(2880)5/2^+$ are
often assigned to $1D$ doublet in the $\Lambda_c$ family~\cite{Chen:2016iyi,Chen:2017aqm,Yao:2018jmc}. It indicates that the mass of the $J^P=3/2^+$ $D$-wave state should be smaller than that of the $J^P=5/2^+$ state. Then, if assigning $\Lambda_b(6146)$ and $\Lambda_b(6152)$ to
be $J^P=5/2^+$ and $J^P=3/2^+$ $D$-wave states, respectively, one should face a serious problem of mass reverse.
Whether the structure in the $\Lambda_b\pi^+\pi^-$ spectrum correspond to two states or one state should be further
clarified with more observations in future experiments.

\subsection{$\Lambda_b(2S)$}

\begin{table}[htb]
\begin{center}
\caption{ \label{Lamb2S} Partial widths (MeV) of strong decays for the $\lambda$-mode $2S$-wave $\Lambda_b$ baryon.}
%\footnotesize
\begin{tabular}{c|ccccc}
\hline\hline
 \multirow{2}{*}{Decay mode}        &\multicolumn{2}{c}{$\underline{~~~~~~|J^P = \frac{1}{2}^+,0 \rangle~~~~~~}$} \\
             &~~ $\Sigma_b(6146)$         &$\Sigma_b(6152)$     \\ \hline
 $\Sigma_b\pi$                                       &0.34                    &0.32                    \\
 $\Sigma_b^*\pi$                                     &1.72                    &1.77   \\
Sum                                                  &2.06                    &2.09  \\
\hline
\hline
\end{tabular}
\end{center}
\end{table}

In the $\Lambda_b$ family, there is only one $\lambda$-mode $2S$-wave
excitation $|J^P = \frac{1}{2}^+,0 \rangle$ according to the quark model classification.
The mass for the $\lambda$-mode $2S$-wave $\Lambda_b$
excitation is predicted to be $\sim 6.1$ GeV in various quark models (see Table~\ref{sp1}).
According to the predicted masses in Ref.~\cite{Yoshida:2015tia},
the measured mass of the $\Lambda_b(6146)$ or $\Lambda_b(6152)$ indicates that it might be a
good candidate of the $\lambda$-mode $2S$-wave excitation.
Considering $\Lambda_b(6146)$ or $\Lambda_b(6152)$ as the $2S$-wave state,
the strong decay properties are studied, our results are listed in Table \ref{Lamb2S}.
It is found that if assigning $\Lambda_b(6146)$ to the $2S$-wave state,
both the total width $\Gamma\simeq 2$ MeV and the dominant decay mode $\Sigma_b^*\pi$
predicted in theory are consistent with the observations. However, the other
resonance $\Lambda_b(6152)$ cannot be understood in the quark model. It cannot be
assigned to any $D$-wave states in the $\Lambda_b$ family, because the mass splitting between the $2S$-wave
and $D$-wave $\Lambda_b$ states is $\sim50-100$ MeV. The $D$-wave $\Sigma_b$ states should
be excluded as well for their typical mass is $\sim 6.3$ GeV~\cite{Capstick:1986bm,Yao:2018jmc}.
As a whole if we assign $\Lambda_b(6146)$ as the
$\Lambda_b(2S)$ state, the other state $\Lambda_b(6152)$ cannot be reasonably explained according to the classification of quark models and mass splitting~\cite{Yoshida:2015tia}.

Since the $\Lambda_b(2S)$ $|J^P=\frac{1}{2}^+,0 \rangle$ state may not favor $\Lambda_b(6146)$, in this work, we take its mass $M=6045$ MeV as predicted in Ref.~\cite{Capstick:1986bm}, and estimate the strong decay of $\Lambda_b(2S)$  into the $\Sigma_b\pi$ and $\Sigma_b^*\pi$ channels. It is found that $\Lambda_b(2S)$ mainly decays into $\Sigma_b\pi$ and $\Sigma_b^*\pi$ modes. The predicted partial widths are
\begin{eqnarray}\label{1DL}
\Gamma[\Sigma_b\pi ]\simeq 0.21 \ \mathrm{MeV},~~~~~~ \Gamma[\Sigma_b^*\pi ]\simeq 0.39 \ \mathrm{MeV},
\end{eqnarray}
and the corresponding  total decay width reads
\begin{eqnarray}\label{1DL}
\Gamma_{\mathrm{total}}\simeq 0.6 \ \mathrm{MeV}.
\end{eqnarray}
The $\Sigma_b^*\pi$ might be an ideal channel for searching for the $\lambda$-mode $2S$-wave $\Lambda_b$ state $|J^P = \frac{1}{2}^+,0 \rangle$ in future experiments.

\begin{table}[htb]
\begin{center}
\caption{ \label{sigem2S} Partial widths (MeV) of strong decays for the $\lambda$-mode $2S$-wave   $\Sigma_b$ baryons.}
%\footnotesize
\begin{tabular}{c|ccccc}
\hline\hline
 \multirow{2}{*}{Decay mode}        &\multicolumn{1}{c}{$\underline{~~~~~~|J^P = \frac{1}{2}^+,1 \rangle~~~~~~}$} &\multicolumn{1}{c}{$\underline{~~~~~~|J^P = \frac{3}{2}^+,1 \rangle~~~~~~}$} \\
             &~~ $\Sigma_b(6213)$            &~~ $\Sigma_b(6226)$         \\ \hline
 $\Lambda_b\pi$                      &3.82                       &4.72      \\
 $\Sigma_b\pi$                       &$5.32\times10^{-2}$        &$3.63\times10^{-3}$     \\
 $\Sigma_b^*\pi$                     &$6.90\times10^{-2}$        &$1.07\times10^{-2}$      \\
Sum                                  &3.94                        &4.73      \\
\hline
\hline
\end{tabular}
\end{center}
\end{table}

\subsection{$\Sigma_b(2S)$}

We also study the strong decay properties of  two $\lambda$-mode $2S$-wave
excitations $|J^P = \frac{1}{2}^+,1 \rangle$ and $|J^P = \frac{3}{2}^+,1 \rangle$ in the $\Sigma_b$ family according to the quark model classification. The masses for the $\lambda$-mode $2S$-wave $\Sigma_b$
excitations are predicted to be $6.2\sim 6.3 $ GeV in various models (see Table~\ref{sp1}).

To study the strong decay properties of the $2S$ wave $\Sigma_b$ excitations, we adopt the predicted masses in Ref.~\cite{Ebert:2011kk}. Our results are listed in Table \ref{sigem2S}. One can see that $|J^P = \frac{1}{2}^+,1 \rangle$ and $|J^P = \frac{3}{2}^+,1 \rangle$ dominantly decay into the $\Lambda_b\pi$ channel. Their partial widths are predicted to be
\begin{eqnarray}\label{1DL}
\Gamma[\Sigma_b(6213)\rightarrow\Lambda_b\pi ]\simeq 3.82 \ \mathrm{MeV},\\
\Gamma[\Sigma_b(6226)\rightarrow\Lambda_b\pi ]\simeq 4.72 \ \mathrm{MeV}.
\end{eqnarray}
The $\Lambda_b\pi$ might be an ideal channel to look for  these radial excitations in future experiments.

Finally, it should be mentioned that we do not consider the $\Lambda_b(6146)$ and $\Lambda_b(6152)$ as the $2S$ $\Sigma_b$ states,
for the mainly decay  modes and masses of $\Sigma_b(2S)$ predicted in the quark model are inconsistent with the observations.

%and dominantly decay mode

\section{Summary}\label{suma}

In this work, we study the strong decays of the low-lying $\lambda$-mode  $\Lambda_b(1D,2S)$ and $\Sigma_b(2S)$ states. Our results indicate that the newly observed $\Lambda_b(6146)$ might be assigned as the $\Lambda_b(1D)$ $|J^P = \frac{5}{2}^+,2 \rangle$ state, which dominantly decays into $\Sigma_b^*\pi$ channel. The $\Lambda_b(6152)$ seems to favor the $\Lambda_b(1D)$ $|J^P = \frac{3}{2}^+,2 \rangle$ states, with this assignment, its decay behaviors are dominated by the $\Sigma_b\pi$ and $\Sigma_b^*\pi$ channels, which are consistent with the experimental observations. However, if we assign $\Lambda_b(6146)$ and $\Lambda_b(6152)$ to the $J^P=5/2^+$ and $J^P=3/2^+$ $D$-wave states, respectively, one should face a serious problem of mass reverse. Whether the structure in the $\Lambda_b\pi^+\pi^-$ spectrum corresponds to two states or one state should be further clarified with more observations in future experiments. Moreover, we find that the $\Lambda_b(2S)$ $|J^P = \frac{1}{2}^+,0 \rangle$ state mainly decays into $\Sigma_b^*\pi$, and the $\Sigma_b(2S)$ $|J^P = \frac{1}{2}^+,1 \rangle$ and $|J^P = \frac{3}{2}^+,1 \rangle$ dominantly decay into $\Lambda_b\pi$ with narrow widths $\Gamma \simeq 3.82$ MeV and $\Gamma \simeq 4.72$ MeV, respectively. These theoretical predictions may provide helpful information for future experimental searches.

\section*{  Acknowledgments }

This work is supported, in part, by the National Natural Science Foundation of China under Grants No.~11775078,
No.~U1832173, and No.~11705056.
%\appendix
%%%%%%%%%%%%%%%%%%%%%%%%%%%%%%%%%%%%%%%%%%%%%%%%%%%%%%%%%%%%%%%%%%555

\bibliographystyle{unsrt}

\end{document}